\documentclass[runningheads]{llncs}
\usepackage{graphicx}
\usepackage{todonotes}
\usepackage{enumerate,multirow,tabularx,ragged2e,booktabs}

\begin{document}
\title{Improving Query Representations for Dense Retrieval with Pseudo Relevance Feedback: \\A Reproducibility Study}
\titlerunning{Improving QR for DR with PRF: A Reproducibility Study}
%
\author{Hang Li\inst{1} \and
Shengyao Zhuang\inst{1} \and
Ahmed Mourad\inst{1} \and
Xueguang Ma\inst{2} \and
Jimmy Lin\inst{2} \and
Guido Zuccon\inst{1}}
\authorrunning{Li et al.}
%
\institute{The University of Queensland, St. Lucia, Australia
\email{\{hang.li,s.zhuang,a.mourad,g.zuccon\}@uq.edu.au}
 \and
 University of Waterloo, Waterloo, Canada \\
 \email{\{x93ma,jimmylin\}@uwaterloo.ca}}

%
\maketitle              
\begin{abstract}

Pseudo-Relevance Feedback (PRF) utilises the relevance signals from the top-$k$ passages from the first round of retrieval to perform a second round of retrieval aiming to improve search effectiveness. A recent research direction has been the study and development of PRF methods for deep language models based rankers, and in particular in the context of dense retrievers. Dense retrievers, compared to more complex neural rankers, provide a trade off between effectiveness, which is often reduced compared to more complex neural rankers, and query latency, which also is reduced making the retrieval pipeline more efficient. The introduction of PRF methods for dense retrievers has been motivated as an attempt to further improve their effectiveness.

In this paper, we reproduce and study a recent method for PRF with dense retrievers, called ANCE-PRF. This method concatenates the query text and that of the top-$k$ feedback passages to form a new query input, which is then encoded into a dense representation using a newly trained query encoder based on the original dense retriever used for the first round of retrieval. While the method can potentially be applied to any of the existing dense retrievers, prior work has studied it only in the context of the ANCE dense retriever.

We study the reproducibility of ANCE-PRF in terms of both its training (encoding of the PRF signal) and inference (ranking) steps. We further extend the empirical analysis provided in the original work to investigate the effect of the hyper-parameters that govern the training process and the robustness of the method across these different settings. Finally, we contribute a study of the generalisability of the ANCE-PRF method when dense retrievers other than ANCE are used for the first round of retrieval and for encoding the PRF signal.

\keywords{Pseudo Relevance Feedback \and Dense Retrievers \and Query Representations}
\end{abstract}

\section{Introduction}

Pseudo-Relevance Feedback (PRF) is a retrieval technique that assumes that the top-$k$ results from the first round of retrieval are relevant; PRF then uses this signal to improve the query representation for a second round of retrieval (or re-ranking) in a bid to obtained higher search effectiveness. PRF has been extensively studied and applied to bag-of-word retrieval models: representative techniques are Rocchio~\cite{rocchio1971rocchio}, KL expansion~\cite{zhai2001model,lv2014revisiting}, RM3~\cite{lv2009comparative} and other relevance models~\cite{lavrenko2017relevance}. It is well accepted that PRF tends to improve search effectiveness, and strong bag-of-words baselines often rely on PRF (e.g., BM25+RM3 is a typical baseline combination).


Aside from its use with bag-of-words models, PRF has recently been studied in the context of Transformer~\cite{vaswani2017attention} based deep language models like BERT~\cite{devlin2018bert}, RoBERTa~\cite{liu2019roberta}; examples of such Transformer-based rankers include cross-encoder architectures such as monoBERT~\cite{nogueira2019passage}. These deep language models have been shown very effective for ranking, though, compared to bag-of-words methods, they often require substantially more computational power and are characterised by high query latencies. Their effectiveness can be further improved by PRF -- but this is at the cost of even higher query latencies, rendering the use of PRF on top of BERT-based rankers like monoBERT practically unfeasible~\cite{li2021pseudo}. 

Dense retrievers (DRs) have been proposed as alternatives to the expensive BERT-based rankers based on the cross-encoder architecture~\cite{zhan2020repbert,karpukhin2020dense,xiong2020approximate,khattab2020colbert}. DRs also rely on deep language models like BERT; however instead of training a cross-encoder to encode query and document\footnote{In this paper we use `document' and `passage' interchangeably. Our experiments and the methods considered are in the context of the passage retrieval task. However, the methods can generalise to deal with documents, at the net of the development of strategies for managing the often large size of documents compared to passages~\cite{}.} at the same time, it relies on a bi-encoder architecture where queries and documents are encoded separately. This separation in the encoding allows to pre-compute document representations (which is computationally expensive for large collections) at indexing, thus leaving only the encoding of the query and the matching between query and documents representations to be performed at query time. Dense retrievers provide a trade off between effectiveness and efficiency: while they are often less effective than the cross-encoder methods, DRs are more efficient (lower query latency). PRF with DRs then becomes suddenly more interesting than when applied to cross-encoders: PRF could provide effectiveness boosts while the additional computational cost imposed by the feedback, infeasible when considering cross-encoders, may be feasible in the context of DRs. This research direction has therefore attracted increasing interest~\cite{li2021pseudo,yu2021improving,wang2021pseudo}.

In this paper, we consider a specific method for PRF with DRs: the ANCE-PRF method~\cite{yu2021improving}. The method uses the ANCE dense retriever~\cite{xiong2020approximate} to perform a first round of retrieval for a given query. Then, the text of the original query is concatenated with that from the top-$k$ documents retrieved by ANCE. The output is a new text query. Then, this new query is encoded using the purposely trained ANCE-PRF encoder to obtain a new dense query representation that is in turn used for computing the match with the documents dense representations to determine a ranking for the query. The ANCE-PRF encoder is trained using a straightforward training procedure with negative sampling strategy based on the original ANCE model, expect that the input to the ANCE-PRF encoder is the concatenation of the query and the relevance signal (top-$k$ documents), rather than just the query (or just the document) as in ANCE.

Given the ANCE-PRF method, we aim to replicate the initial study by Yu et al.~\cite{yu2021improving} in terms of both the training of the ANCE-PRF encoder and its use for retrieval. In addition, we also aim to further extend that work by considering the factors that affect the training of the ANCE-PRF encoder, i.e., the hyper-parameters of the model, and studying their effect on the model performance and therefore its robustness across hyper-parameters settings. We also study the generalisability of the strategy underlying ANCE-PRF to other DRs. In doing so, we develop and publicly release a codebase that implements Yu et al.'s method, along with trained checkpoints of the method for both ANCE and other DRs.

\section{Related Work}
\vspace{-10pt}
Pseudo-Relevance Feedback (PRF) is a classic query expansion method that aims to mitigate the mismatch between query intent and query representation~\cite{clinchant2013theoretical,wang2020pseudo}, by modifying the original query with top-$k$ initially retrieved results. Typical PRF approaches such as Rocchio~\cite{rocchio1971rocchio}, query-regularized mixture model~\cite{tao2006regularized}, KL expansion~\cite{lv2014revisiting,zhai2001model}, RM3\cite{lv2009comparative}, relevance models~\cite{lavrenko2001relevance}, and relevance-feedback matrix factorization~\cite{zamani2016pseudo} are well studied. However, most of the existing studies of PRF methods are applied on top of bag-of-words retrieval models.

With the emerging of transformer~\cite{vaswani2017attention} based models, many researchers have been looking into how to integrate PRF with deep language models. Zheng et al.~\cite{zheng-etal-2020-bert} presented a BERT\cite{devlin2018bert}-based PRF model: BERT-QE, which splits the PRF documents into smaller chunks and utilises the BERT model to identify the most relevant PRF document chunks and use these chunks as PRF signals. Li et al.~\cite{li2018nprf} proposed a neural PRF approach that uses a feed-forward neural network model to aggregate the query and feedback document relevance scores and provide the target document's relevance score. Yu et al.~\cite{yu2021pgt} utilises graph transformers to capture the PRF signals from the initial retrieved results, and Wang et al.~\cite{wang2020pseudo} proposed a clustering method to gather the relevance signals from PRF documents. These methods show remarkable improvements, but the efficiency is significantly affected, such as BERT-QE inference requires 11.01x more computations than BERT alone, making these models computationally infeasible for many practical applications.

Recently, \textit{dense retrievers}~\cite{xiong2020approximate,lin2020distilling,karpukhin2020dense,khattab2020colbert,hofstatter2021efficiently} have been attracting a lot of attention from researchers. These models, which often utilise a BERT-based dual-encoder to encode queries and passages into a shared embedding space, have shown great effectiveness and efficiency in various tasks and datasets. However, most of the existing studies are focusing on different training methods, especially negative sampling techniques~\cite{xiong2020approximate,gao2021complement,lee2019latent}. Most of these models encode either the query or the document to a single embedding vector~\cite{xiong2020approximate,lin2020distilling,hofstatter2021efficiently}, which fits perfectly to many vector-based PRF methods.

In recent research, because of the nature of dense retrievers that uses embedding vectors to represent query and document, different methods have been studied to integrate pseudo relevance information into dense retrievers. Li et al.~\cite{li2021pseudo} investigated two simple approaches, Average and Rocchio, to utilise PRF information in dense retrievers (ANCE~\cite{xiong2020approximate} and RepBERT~\cite{zhan2020repbert}) without introducing new neural models or further training. According to the results, both models achieved superior effectiveness with these two simple approaches without hurting the efficiency significantly compared to the original models, which shows the viability of integrating PRF signals in deep language models. Instead of the simple approaches, researchers have been trying to utilise the pattern learning ability of transformer models to leverage the PRF signals. A recent work from Yu et al.~\cite{yu2021improving}  replaced the query encoder in ANCE~\cite{xiong2020approximate} model by training a new query encoder, which takes the original query text and the PRF documents text together as the new query, based on the original ANCE model as the initial training checkpoint, without changing the document encoder. However, it has several major limitations: 1) for each different PRF depths, it requires to train a new query encoder; 2) the input length for the query encoder is limited, which means the PRF depth is limited; 3) the new query encoder is trained upon ANCE query encoder, which means for different datasets, different ANCE models need to be trained first, making this new approach hard to be generalised.

\section{Improving Query Representations for Dense Retrievers with Pseudo Relevance Feedback}

Next we briefly describe the ANCE-PRF method~\cite{yu2021improving}, which extends  ANCE~\cite{xiong2020approximate} to accept PRF signal from the top-$k$ documents to be encoded in combination with the query to form a new query representation.


In ANCE, the score of a document $d$ for a query $q$ is computed by separately encoding $q$ and $d$ using the ROBERTa~\cite{liu2019roberta} pre-trained deep language model, and then calculating the inner product between the resulting dense representations:


\begin{equation}
	\label{eq:ance-encoder}
	f_{\mbox{ANCE}}(q,d) = \mbox{ANCE}^{\mbox{q}}(\langle s \rangle q \langle /s \rangle) \cdot \mbox{ANCE}^{\mbox{d}}(\langle s \rangle d \langle /s \rangle)
\end{equation}

\noindent where $\mbox{ANCE}^{\mbox{q}}$ and $\mbox{ANCE}^{\mbox{d}}$ represent the query and the document encoders, respectively, and $\langle s \rangle$ and $\langle /s \rangle$ represent the \texttt{[CLS]} and \texttt{[SEP]} tokens in ANCE. Both encoders use the final layer of the $\langle s \rangle$ token embedding as the query and document dense representations.
In ANCE, the document embeddings are pre-computed offline and stored in an index, while the query embeddings are encoded at inference (query) time~\cite{xiong2020approximate}. For fine-tuning Equation~\ref{eq:ance-encoder}, ANCE adopts noisy contrastive estimation loss and employs a negative sampling strategy where negative samples are dynamically retrieved from an asynchronously updated ANCE document index~\cite{xiong2020approximate}.


ANCE-PRF uses a similar schema to score documents for retrieval: 

\begin{equation}
	\label{eq:anceprf-encoder}
	f_{\mbox{ANCE-PRF}}(q,d) = \mbox{ANCE}^{\mbox{prf}}(\langle s \rangle q \langle /s \rangle d_1 \langle /s \rangle ... d_k \langle /s \rangle)  \cdot \mbox{ANCE}^{\mbox{d}}(\langle s \rangle d \langle /s \rangle)
\end{equation}

\noindent where $\mbox{ANCE}^{\mbox{prf}}$ is the newly trained PRF query encoder and $\langle s \rangle q \langle /s \rangle d_1 \langle /s \rangle ...$   $d_k \langle /s \rangle$ is the text concatenation of the original query $q$ with the feedback documents $d_1, d_2,...,d_k$ (in addition to CLS and separator tokens). We denote $q^{prf}$ as the query embedding generated through PRF by $\mbox{ANCE}^{\mbox{prf}}$. 

For the training of the PRF query encoder $ANCE^{prf}$, ANCE-PRF uses the standard noisy contrastive estimation loss:

\begin{equation}
	\label{eq:loss}
	\mathcal{L} = -log\frac{exp(q^{prf} \cdot d^{+})}{exp(q^{prf} \cdot d^{+}) + \Sigma_{d^{-} \in D^{-}}exp(q^{prf} \cdot d^{-})}
\end{equation}

\noindent where $d^+$ represents a relevant document for the query, $d^-$ represents an irrelevant document  (obtained from the negative sampling technique). During the training process, the ANCE-PRF model uses the document embeddings from the original ANCE model. Therefore, the document embeddings remain unchanged in the ANCE-PRF model: it is only the query embedding that changes into the PRF query embedding $q^{prf}$.

Intuitively, ANCE-PRF should provide increases in search effectiveness because the newly trained ANCE-PRF query encoder learns to extract relevant information for the query from the PRF documents using the Transformer attention mechanism~\cite{vaswani2017attention}. After training, the ANCE-PRF query encoder would then pay more attention to the relevant tokens in the PRF documents, while ignoring the irrelevant tokens from this signal. Although Yu et al.~\cite{yu2021improving} do not report the query latency of ANCE-PRF, this should be approximately twice that of the original ANCE model.

\section{Experimental Settings}

\subsection{Datasets}
\vspace{-5pt}
The datasets used in the original work of Yu et al.~\cite{yu2021improving} are TERC DL 2019~\cite{craswell2020overview}, TREC DL 2020~\cite{craswell2021overview}, DL Hard~\cite{mackie2021dlhard}, and MS MARCO Passage Ranking V1~\cite{nguyen2016ms}. These datasets are based on the same corpus provided by MS MARCO Passage Ranking V1, which has $\sim8.8$M passages in total. Note that for TREC DL 2019/2020 queries, each query has multiple judgements on a relevance scale from 0 to 3, while MS MARCO Passage Ranking V1 only has an average of 1 judgement per query with binary relevance, either 0 or 1.

%

The original paper used the training split from MS MARCO Passage Ranking V1 for training ANCE-PRF, which includes $\sim530$K queries. The trained models are evaluated on TREC DL 2019 (43 judged queries), DL 2020 (54 judged queries), DL HARD, and MS MARCO Passage Ranking V1 Dev set (6,980 queries).  For direct comparison with the ANCE-PRF model, we follow the same process except for evaluation on TREC DL HARD (the results on this dataset for other dense retrievers considered in this paper are not publicly available).

\subsection{Models}

The original work by Yu et al.~\cite{yu2021improving} only considers ANCE as the initial dense retriever. To validate their hypothesis that their PRF method can be generalised to other dense retrievers, we consider two recently published dense retrievers that achieve higher performance than ANCE, TCT ColBERT V2 HN+~\cite{lin2020distilling} and DistilBERT KD TABS~\cite{hofstatter2021efficiently}. TCT ColBERT V2 HN+ uses a BERT-style encoder, as shown in Equation~\ref{eq:tct-encoder}, to encode queries and documents, while DistilBERT KD TABS uses a DistilBERT style encoder, as shown in Equation~\ref{eq:dbert-encoder}. These two dense retrievers are different from ANCE with respect to the training process, and the inference is slightly different from each other. We refer the reader to the original papers for further details. The output of these three dense retrievers are all embedding vectors that represent either the query or the document based on the input. The index for all three models are pre-computed and stored offline.

\begin{equation}
    \label{eq:tct-encoder}
	q^{prf}_{TCT} = TCT^{prf}([CLS]\:[Q]\:q[SEP] d_1 [SEP] ... d_k [MASK]*512)
\end{equation}

The scoring function for TCT ColBERT V2 HN+ is:
\begin{equation}
	\label{eq:tct-ret}
	f_{\mbox{TCT-PRF}}(q,d) = q^{prf}_{TCT} \cdot TCT^{d}([CLS]\:[D]\:d)
\end{equation}

where $TCT^{prf}$ represents the new PRF query encoder based on TCT ColBERT V2 HN+ query encoder, the input need a \texttt{[CLS]} token as well as a \texttt{[Q]} in text as prepend to the actual query text, then the PRF document texts are separated by \texttt{[SEP]} token, then pad \texttt{[MASK]} token to fill in the gap if the input is smaller than the max input size of the model, which is 512 for BERT-based models~\cite{devlin2018bert}.

For retrieval, as in Equation~\ref{eq:tct-ret}, $TCT^d$ represents the document encoder, and the input document text is prepended with \texttt{[CLS]} token and \texttt{[D]} in text.

For DistilBERT KD TABS, Equation~\ref{eq:anceprf-encoder} becomes:

\begin{equation}
    \label{eq:dbert-encoder}
	q^{prf}_{DBERT} = DBERT^{prf}([CLS] q [SEP] d_1 [SEP] ... d_k [SEP])
\end{equation}

And the scoring function for DistilBERT KD TABS is:

\begin{equation}
	f_{\mbox{DBERT-PRF}}(q,d) = q^{prf}_{DBERT} \cdot DBERT^{d}([CLS]d[SEP])
\end{equation}

Similar to TCT ColBERT V2 HN+, except the input is a standard BERT input with \texttt{[CLS]} token as prepend and \texttt{[SEP]} token as separators to separate the PRF documents for both PRF query encoding and retrieval.

\subsection{Inferencing and Training}

\subsubsection{Inferencing} To reproduce the ANCE-PRF results, the authors have provided us with a model checkpoint of PRF depth 3. Since there is no inference code available from the original authors, we utilise the open source IR toolkit Pyserini\footnote{https://github.com/castorini/pyserini}~\cite{lin2021pyserini}, which has already implemented the ANCE dense retriever, by introducing a second round of ANCE retrieval with the ANCE-PRF model checkpoint. During the inference time, the document index is the same for both the first round ANCE retrieval and the second round ANCE-PRF retrieval. The only difference in this process is the initial retrieval uses the ANCE query encoder, while the second retrieval uses the ANCE-PRF query encoder.

\subsubsection{Training} Apart from the inference, the authors have not released the training code as well. To reproduce the ANCE-PRF training process, we utilise the open source dense retriever training toolkit Tevatron\footnote{https://github.com/texttron/tevatron}. According to the original paper~\cite{yu2021improving}, all hyperparameters used in ANCE-PRF training are the same as ANCE training, and the ANCE-PRF query encoder is initialised from ANCE FirstP model\footnote{https://github.com/microsoft/ANCE}~\cite{xiong2020approximate}. Although some of the parameters are still not clear, we tried our best to replicate the same model as ANCE-PRF with $k=3$ by adjusting different training settings. 

We also experimented with two stronger dense retrievers, TCT ColBERT V2 HN+~\cite{lin2020distilling} and DistilBERT KD TASB~\cite{hofstatter2021efficiently}, with the same training process to investigate the generalisability of the ANCE-PRF model. Therefore, we adopted the same hyperparameters from these two models and trained with the same settings as ANCE-PRF.

All models in our experiments are trained on two Tesla V100 SMX2 32GB GPUs. In the original paper, the ANCE-PRF model is trained with per device batch size 4 and gradient accumulation step 8 for 450K steps, which is equivalent to per device batch size 32 for $\sim$56K steps, therefore, in our training experiments, we use 10 epochs which is roughly $\sim$80K steps.
\vspace{-3pt}
\subsection{Evaluation Metrics}
\vspace{-3pt}
The official evaluation metric for MS MARCO Passage Ranking V1 dataset is MRR@10~\cite{nguyen2016ms}, for TREC DL 2019 and 2020 are nDCG@10, Recall@1000~\cite{craswell2020overview,craswell2021overview}. For the Recall@1000 evaluation metric on TREC DL 2019 and 2020, the judgements are binarized at relevance point 2 according to the official guideline. Besides the official evaluation metrics, the authors in the original work~\cite{yu2021improving} also use HOLE@10 as an additional evaluation metric to measure the unjudged fraction of top 10 retrieved documents~\cite{xiong2020approximate}, in the reflection of the coverage of the pooled labels on these dense retrieval systems. However, in our experiments, we are going to keep the official evaluation metrics only, for the sake of comparison with other models and baselines. Statistical significant difference between model results are tested using two tails paired t-test. 

\subsection{Research Questions}

In this work, we aim to address the following research questions along with the reproduction of the original method from Yu et al.~\cite{yu2021improving}:

\begin{enumerate}[\indent(a)]
	\item[\textbf{RQ1:}] What is the possibility of replicating the inference results of a ANCE-PRF given only a checkpoint of the trained model provided by the original authors?
	
	\item[\textbf{RQ2:}] The training process is governed by a number of hyper-parameters and choices, importantly including learning rate, optimizer, and negative sampling technique settings. Given the insufficient details in the original study, it is reasonable to expect that researchers attempting to reproduce the ANCE-PRF method may set these parameters to  values different from those in the original study. We are then interested to study: what is the impact of ANCE-PRF training hyper-parameters on the effectiveness of the method, and in particular if this is robust to different hyper-parameter settings?
	\item[\textbf{RQ3:}] The PRF strategy underlying ANCE-PRF can be adapted to other dense retrievers as observed by Yu et al.~\cite{yu2021improving}, but not empirically validated. The original ANCE-PRF model is only trained with ANCE~\cite{xiong2020approximate} as the initial dense retriever. We are then interested to investigate: do the improvements observed for ANCE-PRF generalise to other dense retrievers, such as the two more effective models, TCT ColBERT V2 HP+~\cite{lin2020distilling} and DistilBERT KD TASB~\cite{hofstatter2021efficiently}? 
\end{enumerate}

\section{Results and Analysis}

\subsection{RQ1: Reproduce ANCE-PRF Inference}

A benefit of the transformer-based neural models nowadays is its easy reproducibility. The results can be replicated easily by using the model checkpoint. Therefore, with the PRF 3 checkpoint provided by the authors, we tried to reproduce the same results reported in the original paper, the outcomes are shown in Table~\ref{table:repo-inference}. During the replication process, we found that the ANCE-PRF model is sensitive to the uppercase or lowercase of the letters. For the original queries used in all three datasets in this experiment, no uppercase letters existed, therefore this detail is omitted from the paper. But from our replication experiments, uppercase letters exists in the collection corpus, and the token ids and their associated tokens embeddings are different with different case of the same word. Therefore, for PRF queries, after concatenate the PRF documents with the original query text, the new PRF queries contain uppercase letters and leads to different tokens after tokenization, and resulting in different performance compares to what reported in the original paper. On the other hand, if we set the tokenizer to do lowercase at inference time, then we can get the results same as the original paper. Hence, we successfully reproduced the ANCE-PRF model for inferencing by using the checkpoint provided by the authors.

\begin{table}[]
\centering
\caption{The reproduced results with the provided ANCE-PRF 3 checkpoint after inference. \textbf{unhandled} represents results with the PRF query contains both uppercase and lowercase letters. \textbf{do lower case} indicates the results with  PRF query converted to lowercase when tokenize. \textbf{ANCE-PRF 3} shows results in original paper.}
\resizebox{\columnwidth}{!}{%
\begin{tabular}{l|ccc|cc|cc}
\toprule
Datasets & \multicolumn{3}{c|}{MS MARCO} & \multicolumn{2}{c|}{TREC DL 2019} & \multicolumn{2}{c}{TREC DL 2020} \\ \midrule
              & MRR@10 & nDCG@10 & R@1000 & nDCG@10 & R@1000 & nDCG@10 & R@1000 \\ \midrule
ANCE~\cite{xiong2020approximate} & 0.330 &  0.388 & 0.959 & 	0.648 & 0.755 & 0.646 & 0.776 \\
ANCE-PRF 3~\cite{yu2021improving} & 0.344 & 0.401 & 0.959 & 0.681 & 0.791 & 0.695 & 0.815 \\ \midrule
unhandled     & 0.342 & 0.399  & 0.960 & 0.678  & 0.792 & 0.674  & 0.794 \\
do lower case & 0.344 & 0.402  & 0.960 & 0.681  & 0.791 & 0.695  & 0.815 \\ \bottomrule
\end{tabular}}
\label{table:repo-inference}
\end{table}

To answer RQ1, we confirmed that it is possible to reproduce the same results with the model checkpoint, however one key detail that was missing in the paper is the lower case process to the PRF query. We make our ANCE-PRF inference implementation publicly available in Pyserini toolkit~\footnote{https://github.com/castorini/pyserini/blob/master/docs/experiments-ance-prf.md} so that practitioners can easily reproduce the same results in the original paper with the author provided model checkpoint.

\subsection{RQ2: Reproduce ANCE-PRF Training}

Apart from the inferencing, which is fairly easy by using the checkpoint provided by the original authors, we would like to see if we can reproduce the model by following the training settings provided in the paper. However, some details were missing and we had to consult with the original authors to identify their exact settings. After clarifying the training parameters, we used the same setting to train our own ANCE-PRF model, the results are shown in Table~\ref{table:repo-train}.

\begin{table}[]
\centering
\caption{The reproduced results with the trained ANCE-PRF 3 checkpoint after inference. \textbf{ANCE-PRF 3} shows the results from the original paper. \textbf{ANCE} represents the results from the original ANCE model. \textbf{Reproduced} is the results from our reproduced ANCE-PRF model.}
\resizebox{\columnwidth}{!}{%
\begin{tabular}{l|ccc|cc|cc}
\toprule
Datasets & \multicolumn{3}{c|}{MS MARCO} & \multicolumn{2}{c|}{TREC DL 2019} & \multicolumn{2}{c}{TREC DL 2020} \\ \midrule
              & MRR@10 & nDCG@10 & R@1000 & nDCG@10 & R@1000 & nDCG@10 & R@1000 \\ \midrule
ANCE~\cite{xiong2020approximate} & 0.330 &  0.388 & 0.959 & 	0.648 & 0.755 & 0.646 & 0.776 \\
ANCE-PRF 3~\cite{yu2021improving} & 0.344 & 0.401 & 0.959 & 0.681 & 0.791 & 0.695 & 0.815 \\ \midrule
Reproduced     & 0.347 & 0.405 & 0.963 & 0.672 & 0.794 & 0.701 & 0.814 \\ \bottomrule
\end{tabular}}
\label{table:repo-train}
\end{table}

From the results, once their setting was replicated, we obtained results that are close to those reported and with similar trends, at time worse, other times better, but never statistically significantly different between the reproduced results and the results reported in the original paper. The minor differences between the two results can be potentially explained by random neuron drop out during training and the random seed while sampling the hard negatives from the initially retrieved ANCE results. 



In the original study, the authors reported that they were using all hyper-parameters from ANCE~\cite{xiong2020approximate} training, and all models are trained on two RTX 2080 Ti GPUs with per-GPU batch size 4 and gradient accumulation step 8 for 450K steps. However, some parameters are still unclear in ANCE training. We trained the ANCE-PRF model with two Tesla-V100 SMX2 32GB GPUs with per-GPU batch size 32, learning rate $1e-5$, no in batch negatives or cross batch negatives, and no gradient accumulation steps for 10 epoches. The reason why we choose to remove the gradient accumulation step setting is because we are using GPUs with larger memory. In the original settings, 450K steps with gradient accumulation step 8 and per-GPU batch size 4 is the same as 56,250 steps with per-GPU batch size 32. Therefore, in our training process, we use 10 training epoches, which is equivalent to 83,240 steps in total, and it is already more than the steps used in the original settings.

The optimizer in the training process for the ANCE-PRF model reported in the original study is LAMB optimizer, which we did not notice at first, instead we used the AdamW optimizer which might lead to unsuccessful replication.

A common practice for training new models based on another model is to re-initialise the linear head layer and train from scratch while keep the model body, which is exactly what we have done at first, but it appears that ANCE-PRF model is trained with everything inherited from the ANCE model, including the embedding head and normalisation (linear head layer), so without keeping the linear head layer from ANCE, our trained ANCE-PRF is significantly worse than the original ANCE-PRF model, as shown in Table~\ref{table:repo-pooler}.

\begin{table}[]
\centering
\caption{\textbf{Initial} represents the results by re-initialising the linear head layer. \textbf{Inherit} represents the results by inheriting the linear head layer from ANCE. \textbf{In Batch} represents the results by using in batch negatives. \textbf{No In Batch} represents the results by not using in batch negatives. \textbf{1e-6} represents the results by using 1e-6 as the learning rate. \textbf{1e-5} represents the results by using 1e-5 as learning rate.}
\resizebox{0.6\columnwidth}{!}{%
\begin{tabular}{cccc}
\toprule
Dataset & MS MARCO & \multicolumn{2}{c}{TREC DL 2019} \\ \midrule
        & MRR@10   & nDCG@10         & R@1000         \\ \midrule
Initial   & 0.313   & 0.631          & 0.710         \\
Inherit    & 0.335   & 0.680          & 0.798         \\ \midrule
In Batch   & 0.347   & 0.672          & 0.797         \\
No In Batch    & 0.347   & 0.673          & 0.794         \\ \midrule
1e-6   & 0.335   & 0.680          & 0.798         \\
1e-5    & 0.347   & 0.673          & 0.794         \\ \bottomrule
\end{tabular}}
\label{table:repo-pooler}
\end{table}

According to recent superior models such as RocketQA~\cite{qu2021rocketqa}, uniCOIL~\cite{lin2021few}, show that in-batch negatives helps the model to learn and achieves better performance. However, in our experiments, in-batch negatives does not help to improve the model performance, as shown in Table~\ref{table:repo-pooler}, the difference between using in-batch negatives and without using in-batch negatives are not significant.

Learning rate also plays an important part in the training process, we have experimented with two different learning rates, 1e-5 and 1e-6, the results are shown in Table~\ref{table:repo-pooler}. By using larger learning rate, it tends to improve the MRR@10 for MS MARCO dataset, with smaller learning rate, it tends to improve nDCG@10 and R@1000 in TREC DL 2019. However, only the MRR@10 difference is significant.

In addition to the settings above, we also experimented with different number of hard negatives from the ANCE initially retrieved results. More specifically, we tried to sample 2, 8, and 21 negative samples from the top 200 or top 1000 results, the outcome meets our expectation, with 21 negative samples from top 200 gives the best performance.

To answer RQ2, some hyperparameters, such as learning rate, number of negatives, and optimizer, are crucial for reproducing the model checkpoint.

%
%

\subsection{RQ3: Generalisability of ANCE-PRF Beyond ANCE}

After successfully replicating the ANCE-PRF model both in inference and training, we investigate if this PRF strategy integrated with other popular and effective dense retrievers provide some improvements in effectiveness when compared to the dense retrievers results without PRF. However, this improvement is of a smaller magnitude than that observed for ANCE, which can be observed from Table~\ref{table:general}. This may be due to (1) The best hyper-parameter settings for ANCE-PRF may not be adequate to generalise to other dense retrievers, and different settings may lead other dense retrievers to obtain larger improvements; this speaks to the limited robustness of ANCE-PRF's training strategy. (2) The dense retrievers we consider, TCT ColBERT V2 HN+\cite{lin2020distilling} and DistilBERT KD TASB~\cite{hofstatter2021efficiently} are more effective than ANCE. The limited improvement then may be due to the fact that it is easier to improve a weaker model (ANCE) than it is to improve a more effective one (TCT ColBERT V2 HN+ and DistilBERT KD TASB). 

\begin{table}[]
\caption{Results by training two stronger dense retrievers with the same training process as ANCE-PRF. for direct comparison, we trained model with PRF depth 3.}
\resizebox{\columnwidth}{!}{%
\begin{tabular}{l|ccc|cc|cc}
\toprule
Datasets & \multicolumn{3}{c|}{MS MARCO} & \multicolumn{2}{c|}{TREC DL 2019} & \multicolumn{2}{c}{TREC DL 2020} \\ \midrule
                         & MRR@10 & nDCG@10 & R@1000 & nDCG@10 & R@1000 & nDCG@10 & R@1000 \\ \midrule
ANCE~\cite{xiong2020approximate} & 0.330 &  0.388 & \textbf{0.959} & 	0.648 & 0.755 & 0.646 & 0.776 \\
ANCE-PRF 3~\cite{yu2021improving} & \textbf{0.344} & \textbf{0.401} & \textbf{0.959} & \textbf{0.681} & \textbf{0.791} & \textbf{0.695} & \textbf{0.815} \\ \midrule
TCT ColBERT V2 HN+       & \textbf{0.359} & \textbf{0.420}  & 0.970 & 0.720  & 0.826 & 0.688  & \textbf{0.843} \\
TCT ColBERT V2 HN+ PRF 3 & 0.357 & 0.418  & \textbf{0.971}$^\dagger$ & \textbf{0.741}  & \textbf{0.852}$^\dagger$ & \textbf{0.712}$^\dagger$  & 0.840 \\
DistilBERT KD TASB       & 0.344 & 0.407  & \textbf{0.977} & 0.721  & 0.841 & 0.685  & \textbf{0.873} \\
DistilBERT KD TASB PRF 3 & \textbf{0.348} & \textbf{0.411}$^\dagger$  & 0.974 & \textbf{0.736}  & \textbf{0.857}$^\dagger$ & \textbf{0.698}$^\dagger$  & 0.866 \\ \bottomrule
\end{tabular}}
\label{table:general}
\end{table}

To answer RQ3, we find that applying the same training strategy as ANCE-PRF to other more effective dense retrievers only provide smaller magnitude improvement. Hence, the ANCE-PRF method may not generalize to other dense retrievers or requires specific hyper-parameter tuning.

\section{Conclusion}

In this paper we considered the ANCE-PRF model proposed by Yu et al.~\cite{yu2021improving}. This method is the first of its kind to integrate PRF signals directly into the query encoder, without changing the document encoder or document index. 

There are three research questions related to reproducing ANCE-PRF. RQ1 is aimed to address the issues when reproducing the inference results by directly adopting the model checkpoint provided by the original authors. Our experiments show that ANCE-PRF is an uncased model, it can only handle lowercase letters for queries. If the query contains uppercase letters, the ANCE-PRF model performs differently and hurts its performance.

RQ2 is aimed to reproduce the training process of the ANCE-PRF model by using the settings provided in the original study. However, some details are missing which leads to unsatisfying performance of the reproduced model. After consulting with the original authors and using the exact same training settings, we were able to reproduce the ANCE-PRF model with insignificant differences that might be caused by random seed in negative sampling or model initialisation. We then investigate the effect of hyper-parameters in the ANCE-PRF model training. Since some details are left out in the original study, one would reproduce the model by using common practice. However, in our experiments, we found that some common practice might now work in this case. For example, using the linear head layer from ANCE model to train ANCE-PRF is significantly better than re-initialise the linear head layer. In-batch negatives are proved to be useful for training in many superior models, but in our experiment, there is no significant difference between using in-batch negatives and no in-batch negatives.

RQ3 is aimed to test the generalisability of the training method of the ANCE-PRF model. We use the same parameter settings to train the PRF model on top of TCT ColBERT V2 HN+\cite{lin2020distilling} and DistilBERT KD TASB~\cite{hofstatter2021efficiently}, two more effective dense retrievers compared to ANCE. However the results are mixed, the improvements with PRF is of a smaller magnitude than that observed for ANCE, this may be because the best hyper-parameter settings is not suitable for all dense retrievers; to achieve better performance one may need to adjust the parameters accordingly. Another reason may be because both newly added models are more effective than ANCE, the limited improvements may be because this training method is more suitable to improve a weaker model.

The code to reproduce the training of all the models in this work is made available at \url{https://github.com/ielab/APR}.

%
%
%
\bibliographystyle{splncs04}
\bibliography{references.bib}
\end{document}